\documentclass[showpacs,showkeys,preprintnumbers,amsmath,amssymb,nofootinbib,groupedaddress,superscriptaddress,prc]{revtex4}
\usepackage{graphicx}
\usepackage{dcolumn}
\usepackage{bm}

\def\sss{\mbox{\boldmath $\sigma$}}
\newcommand{\be}{\begin{equation}}
\newcommand{\ee}{\end{equation}}
\newcommand{\bea}{\begin{eqnarray}}
\newcommand{\eea}{\end{eqnarray}}
\newcommand{\nn}{\nonumber}

\newcommand{\Image}{\mathop{\rm Im}\nolimits}

\begin{document}

\title{On Gamow-Teller strength distributions for $\beta\beta$-decaying nuclei within continuum-QRPA}

\author{S.Yu. Igashov}
\email{igashov@theor.mephi.ru}
\affiliation{National Research Nuclear University ``MEPhI", 115409 Moscow, Russia}
\author{Vadim Rodin}
\affiliation{Institute f\"{u}r Theoretische Physik der Universit\"{a}t
T\"{u}bingen, D-72076 T\"{u}bingen, Germany}
\author{Amand Faessler}
\affiliation{Institute f\"{u}r Theoretische Physik der Universit\"{a}t
T\"{u}bingen, D-72076 T\"{u}bingen, Germany}
\author{M.H. Urin}
\affiliation{National Research Nuclear University ``MEPhI", 115409 Moscow, Russia}

\begin{abstract} 
An isospin-selfconsistent pn-continuum-QRPA approach is formulated and applied to describe the 
Gamow-Teller strength distributions for $\beta\beta$-decaying open-shell nuclei. 
The calculation results obtained for the pairs of nuclei $^{76}$Ge-Se, $^{100}$Mo-Ru, $^{116}$Cd-Sn, and 
$^{130}$Te-Xe are compared with available experimental data.
\end{abstract}




\date{\today}

\maketitle

\section{Introduction}

Description of weak interaction processes in nuclei is often a challenge for nuclear structure models. 
Numerous calculations of the nuclear double beta ($\beta\beta$) decay amplitudes 
well illustrate this statement (see, e.g., Refs.~\cite{1}). 
Uncertainties in theoretical calculations of the Gamow-Teller (GT) two-neutrino $\beta\beta$
($2\nu\beta\beta$) decay amplitude $M^{2\nu}_{GT}$ have stimulated experimental 
studies of the GT$^{(\mp)}$ strengths for the transitions to the intermediate $1^{+}$ states virtually excited 
in the decay process (see, e.g. Refs.~\cite{2}-\cite{4}). 

As a double charge-exchange process, $2\nu(0\nu)\beta\beta$-decay is enhanced 
by nucleon pairing which is due to the singlet part of the particle-particle (p-p) 
interaction. 
The discrete quasiboson version of the quasiparticle RPA (pn-dQRPA) which accounts for the nucleon pairing 
is usually applied to calculate the $\beta\beta$-decay amplitudes in open-shell nuclei \cite{1}. 
In spite of differences in the model parametrization of the nuclear mean field and the residual interaction in 
the particle-hole (p-h) and p-p channels, all pn-dQRPA calculations reveal
marked sensitivity of 
the amplitude $M^{2\nu}_{GT}$ to the ratio $g_{pp}$ of the triplet to singlet p-p interaction strengths.  
Physical reasons for such a general feature of all calculations were analyzed in 
Ref.~\cite{41}, and the sensitivity was shown to reflect  
violation of the Wigner spin-isospin SU(4) symmetry in nuclei. 
An identity transformation of the amplitude into the sum of two 
terms was used in Ref.~\cite{41}. 
One term, which is due to 
p-p interaction only, depends linearly on $g_{pp}$ and vanishes at $g_{pp}=1$ where
the SU(4)-symmetry is restored in the p-p sector of a model Hamiltonian.
The second term is a smoother function of $g_{pp}$ at $g_{pp}\sim 1$, 
but exhibits a quadratic dependence on the strength of the mean-field spin-orbit term, 
which is the main source of breaking of the spin-isospin SU(4)-symmetry in nuclei.

Understanding of general properties of the amplitude $M^{2\nu}_{GT}$ helps
to improve the reliability of evaluation of the $\beta\beta$-decay amplitudes.
For a quantitative analysis, we use here an isospin-selfconsistent pn-continuum-QRPA (pn-cQRPA) approach, 
which was initially proposed in Ref. \cite{5} to describe 
different strength distributions in single-open-shell nuclei. 
In the reference the full basis of the single-particle (s-p) 
states was used in the p-h channel along with the Landau-Migdal forces, while the nucleon pairing was described 
within the simplest version of the BCS-model based on a discrete s-p basis. 
A rather old version of the phenomenological isoscalar nuclear mean field 
(including the spin-orbit term) was used in Refs.~\cite{41,5}, as well. 
A further development of the pn-cQRPA approach applied to description of the $\beta\beta$-decay matrix elements 
has been given in Ref.~\cite{6}. 
In this reference realistic (zero-range) forces have also been used in the p-p channel to describe 
the nucleon pairing within the BCS model realized on a rather large discrete+quasidiscrete s-p basis.

In the present work along with the detailed formulation of the pn-cQRPA approach we use a modern version 
of the phenomenological isoscalar mean field (including the spin-orbit term) deduced in Ref.~\cite{7} 
from the isospin-selfconsistent analysis of experimental single-particle spectra in double-closed-shell 
nuclei. The last version of the approach and some applications to the GT observables are briefly described in 
Ref.~\cite{8}.

The paper organized as follows.
In Sect. 2 starting from the coordinate representation of the pn-dQRPA equations, we formulate a pn-cQRPA 
approach. The expressions for different type strength functions are also given. 
A realistic model Hamiltonian and the calculation scheme are described in Sect. 3. 
In Sect. 4 the results of pn-cQRPA calculations of the GT$^{\mp}$ strength functions and $M^{2\nu}_{GT}$ 
amplitudes for the number of nuclei are presented and compared with the available experimental data. 
Conclusions are drawn in Sect. 5.

\section{Versions of the pn-QRPA: discrete and continuum}

\subsection{
pn-dQRPA equations in the coordinate representation}

The first step towards formulation of a pn-cQRPA approach is transformation of the pn-dQRPA equations to the 
coordinate representation (see also Refs.~\cite{5,6,8}).
The energies $\omega_s$ and wave functions $|s,J^\pi M\rangle$ of the states in the isobaric (odd-odd) nucleus are usually 
obtained within the quasiboson version of the pn-dQRPA as a solution of a system of homogeneous equations for the forward 
and backward amplitudes $X^{JLS}_s$ and $Y^{JLS}_s$ related to $\beta^{\mp}$ charge-exchange excitations of 
an even-even parent nucleus. Here, $J$ and $M$ are the total angular momentum and its projection  
and $\pi=(-1)^L$ is the parity of the intermediate states ($L=0,1,2\dots$ and $S=0,1$ are the transferred orbital momentum and spin, respectively).

The pn-dQRPA eigenenergy solutions $\omega_s$ are related to the excitation energies 
$E^{(\mp)}_{x,s}$ measured from the ground-state energy $E_0(Z\pm 1,N\mp 1)$ of 
the corresponding daughter nuclei as:
\be
\omega^{(\mp)}_s=\omega_s\pm(\mu_p-\mu_n)=E^{(\mp)}_{x,s}+Q^{(\mp)}_b.
\label{1}
\ee
Here, $\mu_{p(n)}$ is the chemical potential for the proton (neutron) 
subsystem found from the known BCS equations, 
$Q^{(\mp)}_b={\cal E}_b(Z,N)-{\cal E}_b(Z\pm 1,N\mp 1)$ are the total 
binding-energy differences, $\omega^{(\mp)}_s$ are the excitation energies 
measured from $E_0(Z,N)-\sum_a m_a=-{\cal E}_b(Z,N)$ 
($m_a$ is the nucleon mass).
The energies $\omega^{(\mp)}_s$ are usually described by a model Hamiltonian.

For definiteness, we consider further excitations in the $\beta^-$ channel.
The amplitudes $X^{JLS}_{s,\pi\nu}$ and $Y^{JLS}_{s,\pi\nu}$ are based on the discrete s-p levels $\pi$ and 
$\nu$ for protons and neutrons, respectively (each set $\pi$ and $\nu$ contains the s.p. quantum numbers $n_r$, $l$, $j$). 
The excitations in the $\beta^+$-channel are described within the pn-dQRPA by the same equations with substitution 
$p\leftrightarrow n$ ($\pi\leftrightarrow\nu$).

Following Ref.~\cite{5}, we transform the system of equations for the amplitudes $X$ and $Y$ into the coordinate 
representation. 
Transformation is performed in terms of the four-component radial transition density 
$r^{-2}\varrho^{JLS}_{K,s}(r)$ defined as follows:
\bea
&& \varrho^{JLSM}_{K,s}({\bf r})=\frac{1}{r^2}\varrho^{JLS}_{K,s}(r)T_{JLSM}({\bf n}),\ 
\varrho^{JLS}_{K,s}(r)=\sum\limits_{\pi\nu} \varrho^{JLS}_{K,s,\pi\nu} \, 
\chi_{\pi\nu}^{JLS}(r),\label{defvarrho}\\
&&\left({\begin{array}{c}
\varrho_{1,\pi\nu}\\
\varrho_{2,\pi\nu}\\
\varrho_{3,\pi\nu}\\
\varrho_{4,\pi\nu}
\end{array}}\right)
=
\left({\begin{array}{c}
u_\pi v_\nu X_{\pi\nu}+v_\pi u_\nu Y_{\pi\nu}\\
u_\pi v_\nu Y_{\pi\nu}+v_\pi u_\nu X_{\pi\nu}\\
u_\pi u_\nu X_{\pi\nu}-v_\pi v_\nu Y_{\pi\nu}\\
u_\pi u_\nu Y_{\pi\nu}-v_\pi v_\nu X_{\pi\nu}
\end{array}}\right).
\nn
\eea
where $u,~v$ are the coefficients of Bogolyubov transformation;
$\chi_{\pi\nu}^{JLS}(r)=t^{JLS}_{(\pi)(\nu)}\,\chi_\pi(r)\chi_\nu(r)$ 
with $\sqrt{2J+1}$ $t^{JLS}_{(\pi)(\nu)}=$ $\langle(\pi)\|T_{JLS}\|(\nu)\rangle$ 
being the reduced matrix element of the spin-angular tensor $T_{JLSM}({\bf n}\equiv{\bf r}/r)$; $(\pi)=j_{\pi}, l_{\pi}$ ($(\nu)=j_{\nu},l_{\nu}$) and $r^{-1}\chi_{\pi}(r)$ ($r^{-1}\chi_{\nu}(r)$), 
being the s-p proton (neutron) quantum numbers and radial wave functions, respectively. Hereafter 
we shall sometimes omit the superscript ``$JLS$" and/or the subscript ``$s$" when it does not lead to a confusion. 

According to the definition (\ref{defvarrho}), the elements
$\varrho_1,\varrho_2,\varrho_3,\varrho_4$ can be called the particle-hole, hole-particle, particle-particle and 
hole-hole components of the transition density, respectively.
In particular, the transition matrix element to a state $|s,J^{\pi}M\rangle$ corresponding
to a probing particle-hole operator 
\be
\hat V^{(-)}_{JLSM}=\sum_a V_L(r_a)T_{JLSM}({\bf n}_a)\ \tau^{(-)}_a      \label{probop}
\ee
is determined by the element $\varrho_1$ as:
\be 
\langle s,J^\pi M|\hat V^{(-)}_{JLSM}|0\rangle=\int \varrho^{JLS}_{1,s}(r) V_{L}(r)\, dr.
\ee
Assuming the residual interaction $F^{S}_{K}$ in the p-h ($K=1,2$) and p-p ($K=3,4$) channels
to be momentum-independent, one can represent the system of homogeneous pn-dQRPA equations
(primarily written for the amplitudes $X$ and $Y$) in the equivalent form:
\be
\varrho^{JLS}_{K,s}(r)=
\sum\limits_{K'} \int A^{JLS}_{KK'}(r,r',\omega=\omega_s)\, F^{S}_{K'}(r',r'')\, \varrho^{JLS}_{K',s}(r'')
\, dr'dr''.
\label{rhocoord}
\ee
Here, $(rr')^{-1}F^{S}_{K}(r,r')$ is the radial part of the residual interaction,
the $4\times 4$ matrix $(rr')^{-2}A_{KK'}(r,r',\omega)$ is the radial part of the free two-quasiparticle 
propagator hereafter taken diagonal on $L$, $S$ (the so-called "symmetric" approximation \cite{5}):
\bea
&A_{KK'}(r_1,r_2,\omega)=\sum\limits_{\pi\nu} 
\, \chi_{\pi\nu}(r_1)\chi_{\pi\nu}(r_2)\, A_{KK',\pi\nu}(\omega), \ \ \ \ A_{KK',\pi\nu}=A_{K'K,\pi\nu} ,
\label{propdiscr}\\
& A_{11,\pi\nu}=\displaystyle\frac{u^2_\pi v^2_\nu}{\omega-E_{\pi\nu}}-\frac{u^2_\nu v^2_\pi}{\omega+E_{\pi\nu}},
\ \ A_{33,\pi\nu}=\frac{u^2_\pi u^2_\nu}{\omega-E_{\pi\nu}}-\frac{v^2_\nu v^2_\pi}{\omega+E_{\pi\nu}},
\nonumber\\
& \nonumber\\
&A_{12,\pi\nu}=-A_{34,\pi\nu}=\displaystyle
u_\pi v_\pi v_\nu u_\nu
\Bigl(\frac{1}{\omega-E_{\pi\nu}}-
\frac{1}{\omega+E_{\pi\nu}}\Bigr),
\nonumber\\
&A_{13,\pi\nu}=u_\nu v_\nu \displaystyle(\frac{u^2_\pi}{\omega-E_{\pi\nu}}+\frac{v^2_\pi}{\omega+E_{\pi\nu}}),\ \
A_{14,\pi\nu}=-u_\pi v_\pi (\frac{v^2_\nu}{\omega-E_{\pi\nu}}+\frac{u^2_\nu}{\omega+E_{\pi\nu}})\ ,
\nonumber\\
&\ \nonumber\\
& A_{22,\pi\nu}(\omega)=A_{11,\pi\nu}(-\omega),\ \ A_{44,\pi\nu}(\omega)=A_{33,\pi\nu}(-\omega), \ \ 
A_{23\pi\nu}(\omega)=A_{14,\pi\nu}(-\omega),\ \ A_{24,\pi\nu}(\omega)=A_{13,\pi\nu}(-\omega),
\nonumber
\eea
with $E_{\pi\nu}=E_{\pi}+E_{\nu}$, where $E_{\pi}$($E_{\nu}$) is the proton (neutron)
quasiparticle energy. 
Excitations in the $\beta^{+}$-channel are described within the pn-dQRPA 
by Eqs. (\ref{rhocoord}),(\ref{propdiscr}) with the substitution $p\leftrightarrow n$ ($\pi\leftrightarrow\nu$). 
The relationship $A_{KK',\nu\pi}(\omega)= A_{KK',\pi\nu}(-\omega)$, which follows from Eq. (\ref{propdiscr}), 
allows one to reduce pn-QRPA description 
of excitations in the $\beta^{+}$-channel to pn-QRPA description of excitations in the $\beta^{-}$-channel. 
In particular, 
because of equalities $A_{11,\nu\pi}(\omega)=A_{11,\pi\nu}(-\omega)=A_{22,\pi\nu}(\omega)$, the transition matrix
element to the state $|s,J^\pi M\rangle$ corresponding to a probing operator $\hat V^{(+)}_{JLSM}$
(given by Eq. (\ref{probop}) with substitution $\tau^{(-)}\to \tau^{(+)})$ is determined by the
transition-density element $\varrho_{2}$: $\langle s,J^{\pi}M|\hat V^{(+)}_{JLSM}|0\rangle =
\int \varrho ^{JLS}_{2,s}(r) V_{L}(r) dr$. 
In such a description the operator $\hat V^{(+)}_{JLSM}$ can be considered as the hole-particle one.

\subsection {QRPA effective operators
and strength functions}

The next step towards taking the s-p continuum into account is consideration of the
effective two-quasiparticle propagator (two-quasiparticle Green function) $\tilde A_{KK'}
(r,r',\omega)$, which satisfies a Bethe-Salpeter-type equation~\cite{6}: 
\be
\tilde A^{JLS}_{KK'}(r,r',\omega)=A^{JLS}_{KK'}(r,r',\omega)+
\sum\limits_{K''} \int A^{JLS}_{KK''}(r,r_1,\omega) \, F^{S}_{K''}(r_1,r_2)\, \tilde A^{JLS}_{K''K'}(r_2,r',\omega)
\, dr_1dr_2.
\label{2qpprop}
\ee
Here, the current energy $\omega$ is related to the excitation energy of daughter nuclei: 
$\omega^{(\mp)}=\omega\pm(\mu_p-\mu_n)$ (compare with Eq.(1)).
The spectral decomposition of the effective propagator contains all the necessary information about the pn-QRPA solutions:
\be
\tilde A^{JLS}_{11}(r_1,r_2,\omega)=
\sum_s \frac{\varrho^{JLS}_{1,s}(r_1)\varrho^{JLS}_{1,s}(r_2)}{\omega-\omega_s+i0}-
\sum_s \frac{\varrho^{JLS}_{2,s}(r_1)\varrho^{JLS}_{2,s}(r_2)}{\omega+\omega_s-i0},
\label{eqN8}
\ee
\be
\tilde A^{JLS}_{22}(r_1,r_2,\omega)=\tilde A^{JLS}_{11}(r_1,r_2,-\omega),
\label{propexp}
\ee
\be
\tilde A^{JLS}_{12}(r_1,r_2,\omega)=
\sum_s \frac{\varrho^{JLS}_{1,s}(r_1)\varrho^{JLS}_{2,s}(r_2)}{\omega-\omega_s+i0}-
\sum_s \frac{\varrho^{JLS}_{2,s}(r_1)\varrho^{JLS}_{1,s}(r_2)}{\omega+\omega_s-i0}.
\label{eqN10}
\ee

In practical applications of the pn-QRPA method it is more convenient to use 
four-component effective fields $\tilde V_{K[I]}(r,\omega)$ corresponding to the probing operator $V(r)$ acting in the particle-hole
($I=1$) or hole-particle ($I=2$) channels. The effective fields are defined as follows:
\be
\int \tilde A_{KI}(r,r',\omega)\, V(r')\, dr'= \sum\limits_{K'} \int A_{KK'}(r,r',\omega) \tilde V_{K'[I]}(r',\omega)\, dr'.\label{fefexf}
\ee                                              
The system of equations for the effective operators follows from Eqs. (\ref{2qpprop}),(\ref{fefexf}):
\be
\tilde V^{JLS}_{K[I]}(r,\omega)=V_L(r)\delta_{KI}+
\sum\limits_{K'} \int F^{S}_{K}(r,r')\, A^{JLS}_{KK'}(r',r'',\omega)\tilde V^{JLS}_{K'[I]}(r'',\omega)\, dr'dr''.
\label{eqexf}
\ee
Similar equations are given in Ref. \cite{9}, where they are obtained by the methods of the Migdal's 
finite Fermi-system theory.

By definition, the strength functions corresponding to the probing operators of Eq. (\ref{probop}) are:
\be
S_{JLS}^{(\mp)}(\omega)=\sum\limits_{s} \left | \langle s,J^{\pi} \| \hat V^{(\mp)}_{JLS} \|0\rangle\right |^2
\delta(\omega-\omega_s).
\label{defstrf}                   
\ee
Making use of the spectral expansion Eq. (\ref{propexp}), one can express the strength functions
in terms of the two-quasiparticle effective propagator, or in equivalent terms of the corresponding
effective fields: 
\bea
&S^{(-)}_{JLS}(\omega)=-\frac 1\pi (2J+1)\Image \int V_L(r_1) \tilde A_{11}^{JLS}(r_1,r_2,\omega) V(r_2)\, dr_1dr_2=
\nonumber\\
&\ 
\nonumber\\
&=-\frac 1\pi (2J+1)\Image \sum\limits_{K} \int V_L(r_1) A_{1K}^{JLS}(r_1,r_2,\omega) \tilde V_{K[1]}^{JLS}(r_2,\omega)\, dr_1dr_2.
\label{strf-}\eea
\bea
&S^{(+)}_{JLS}(\omega)=-\frac 1\pi (2J+1)\Image \int V_L(r_1) \tilde A_{22}^{JLS}(r_1,r_2,\omega) V(r_2)\, dr_1dr_2=
\nonumber\\
&\ 
\nonumber\\
&=-\frac 1\pi \Image (2J+1)\sum\limits_{K} \int V_L(r_1) A_{2K}^{JLS}(r_1,r_2,\omega) \tilde V_{K[2]}^{JLS}(r_2,\omega)\, dr_1dr_2.
\label{strf+}              
\eea
Within the pn-QRPA description of the $\beta\beta$-decay amplitudes it is useful to consider a ``non-diagonal" 
strength function~\cite{6}:
\be
S^{(- -)}_{JLS}(\omega)=\sum\limits_{s}  \langle 0' \| \hat V^{(-)}_{JLS} \|s,J^{\pi} \rangle 
\langle s,J^{\pi} \| \hat V^{(-)}_{JLS} \|0\rangle \delta(\omega -\omega_s),
\label{defndstrf}
\ee
where $|0'\rangle$ is the ground state wave function of the final nucleus.
Identifying the pn-QRPA vacuum $|0'\rangle$ with that for $|0\rangle$, one gets the following expressions:
\bea
&S^{(- -)}_{JLS}(\omega)=-\frac 1\pi (2J+1)\Image\int V_L(r_1) \tilde A_{21}^{(JLS)}(r_1,r_2,\omega) V_L(r_2) \, dr_1dr_2=
\nonumber\\
&\ 
\nonumber\\
&=-\frac 1\pi (2J+1)\Image \sum\limits_{K} \int V_L(r_1) A_{2K}^{(JLS)}(r_1,r_2,\omega) \tilde V^{JLS}_{K[1]}(r_2,\omega)\, dr_1dr_2.
\label{ndstrf}
\eea
Bearing in mind applications of the method to describing the GT$^{\pm}$ strength functions
and $2\nu\beta\beta$-decay amplitude $M^{2\nu}_{GT}$, in this paper we limit concrete consideration by the case 
$(JLS)=(101)$ and use for this case the probing GT operators
$\hat G^{\mp}_{M}=\sum_{a}\sigma_{M}(a)\tau^{(\mp)}_{a}$. Here $\sigma_{M}$ are the Pauli spherical matrices, the radial 
part of the corresponding external field of Eq.(\ref{probop}) is
$V_{L=0}(r)={\sqrt{4\pi}}$. The integrated strength functions $S^{(\mp)}_{GT}$ satisfy the Ikeda
sum rule equal $3(N-Z)$.

\subsection{$2\nu\beta\beta$-decay amplitude
}\label{defdecam1}

The nuclear GT$^{-}$ amplitude for the $2\nu\beta\beta$-decay into the ground state $|0'\rangle$ of the final nucleus 
(Z+2,N-2) is given by the expression
\be
M^{2\nu}_{GT}=\sum_s\bar\omega^{-1}_s\langle 0'\|\hat G^-\|s,1^+\rangle
\langle s,1^+\|\hat G^-\|0\rangle,
\label{16}
\ee
where $\bar\omega_s=E_s-\frac{1}{2}(E_0+E_{0'})=E_{x,s}+\frac{1}{2}(Q^{(-)}_b+{Q^{(+)}}'_b)$.
To calculate $M^{2\nu}_{GT}$ within the pn-QRPA, the vacua $|0\rangle$ and $|0'\rangle$ should be identified.
As a result of such identification, one has $\bar{\omega}_s=\frac{1}{2}(\omega^{(-)}_s+{\omega^{(+)}}'_s)\approx\omega_s$ 
in accordance with Eq.(1). 
Therefore, the amplitude (\ref{16}) can be 
directly expressed in terms of the ``non-diagonal" GT$^{-}$ strength function 
of Eqs.~(\ref{defndstrf}), (\ref{ndstrf}):
\be
M^{2\nu}_{GT}=\int\omega^{-1}S^{(-\,-)}_{GT}(\omega)d\omega.
\label{17}
\ee
An alternative expression for $M^{2\nu}_{GT}$ is obtained in terms of the ``non-diagonal" static polarizibility~\cite{6}:
\be
M^{2\nu}_{GT}=-3\sqrt{\pi}\sum\limits_{i}\int A_{2i}^{101}(r,r',\omega=0)\tilde V^{(-)}_{i[1]}(r',\omega=0)\, drdr'.
\label{18}
\ee
As mentioned in Introduction , the amplitude of Eq.~(\ref{16}) can be decomposed in a model-independent way into two terms 
\cite{41}:
\be
M^{2\nu}_{GT}=M'_{GT}+M''_{GT},\ \ \ M''_{GT}=\frac{EWSR^{(-\,-)}_{GT}}{\bar\omega_{GTR}^{2}},
\label{19}
\ee
where $\bar\omega_{GTR}$ is the energy of the GT$^{-}$ giant resonance (GTR), and 
$EWSR^{(-\,-)}_{GT}=\int\omega S^{(-\,-)}_{GT}(\omega)d\omega$ is the ``non-diagonal" energy-weighted sum rule.
This sum rule, which is determined by the p-p interaction only, is mainly responsible for the sensitivity of the amplitude 
$M^{2\nu}_{GT}$ to the ratio $g_{pp}$ of the triplet to singlet strength of the p-p interaction.
The sum rule $EWSR^{(-\,-)}_{GT}$ depends linearly on $g_{pp}$ and vanishes at $g_{pp}=1$ where the SU(4) symmetry 
is restored in the p-p sector of a model Hamiltonian. 
The term $M'_{GT}$ in Eq. (\ref{19}) is a smoother function of $g_{pp}$ at $g_{pp}\sim 1$, but exhibits a quadratic 
dependence of the strength of the mean-field spin-orbit term, 
which is the main source of violation of the spin-isospin 
SU(4)-symmetry in nuclei.

\subsection{Exact treatment of the single-particle continuum in the particle-hole channel}

The above given coordinate representation of the pn-dQRPA equations is fully equivalent
to the version usually formulated in terms of the $X$ and $Y$ amplitudes, provided a discrete basis of 
s-p states is used. 
However, the use of the coordinate representation 
allows us to exactly take into account the s-p continuum in the p-h channel and, therefore, to formulate a version of the 
continuum-pn-QRPA (pn-cQRPA) (see also Refs.~\cite{5,6,8}). 
Within this version the pairing problem is solved on a basis of bound and quasibound proton and 
neutron s-p states. To take the s-p continuum into account , the following transformations in Eq. (\ref{propdiscr}) is performed:
(i) the Bogolyubov coefficients $v_{\lambda}$, $u_{\lambda}$ and the quasiparticle energies $E_{\lambda}$ are approximated 
by their non-pairing values $v_{\lambda}$=0, 
$u_{\lambda}=1$ and
$E_{\lambda}=\varepsilon_{\lambda}-\mu$ for those s-p states, which lie far above the chemical potential (i.e., $(\varepsilon_{\lambda}-\mu) \gg \Delta$);
(ii) the radial s-p Green function $g_{(\lambda)}(r,r',\varepsilon)= 
\sum_{\epsilon_{\lambda}}(\epsilon-\epsilon_{\lambda}+i0)^{-1}\chi_{\lambda}(r)\chi_{\lambda}(r')$ is used to explicitly perform the sum over s-p states in the continuum 
($\lambda=\varepsilon_{\lambda},(\lambda)$; $(\lambda)= j_{\lambda},l_{\lambda}$).
As a result, we get from Eq. (\ref{propdiscr}) the following representation of the properly transformed free 
two-quasiparticle propagator: 
\bea
A_{11}(r_1,r_2,\omega)&=&\sum\limits_{\nu_<,\pi_<}
\displaystyle\frac{v^2_\nu u^2_\pi\chi_{\pi\nu}(r_1)\chi_{\pi\nu}(r_2)}{\omega-E_{\pi\nu}}
+\sum\limits_{\nu_<,(\pi)}t^2_{(\pi)(\nu)}\chi_\nu(r_1)\chi_\nu(r_2)~v^2_\nu
g'_{(\pi)}(r_1,r_2,\mu_\pi+\omega-E_\nu)\nonumber\\
&+&\left \{p\leftrightarrow n, \pi\leftrightarrow\nu,\omega\leftrightarrow -\omega \right \},\nonumber\\
&
\nonumber\\
A_{12}(r_1,r_2,\omega)&=&\sum\limits_{\nu_<,\pi_<}
u_\nu v_\nu u_\pi v_\pi\chi_{\pi\nu}(r_1)\chi_{\pi\nu}(r_2)
\Bigl(\displaystyle\frac{1}{\omega-E_{\pi\nu}}-\frac{1}{\omega+E_{\pi\nu}}\Bigr),\nonumber\\
&
\nonumber\\
A_{13}(r_1,r_2,\omega)&=&\sum\limits_{\pi<,\nu}\chi_{\pi\nu}(r_1)\chi_{\pi\nu}(r_2)u_{\nu}v_{\nu}
\Bigl(\displaystyle\frac{u^2_{\pi}}{\omega-E_{\pi\nu}}+\frac{v^2_{\pi}}{\omega+E_{\pi\nu}}\Bigr)\nonumber\\
&
\nonumber\\
&+&\sum\limits_{(\pi),\nu}t^2_{(\pi)(\nu)}\chi_{\nu}(r_1)\chi_{\nu}(r_2)u_{\nu}v_{\nu}g'_{(\pi)}(r_1,r_2,\omega+\mu_p-E_{\nu}),
\nonumber\\
&
\nonumber
\eea

\bea
A_{14}(r_1,r_2,\omega)&=&-\sum\limits_{\pi,\nu<}\chi_{\pi\nu}(r_1)\chi_{\pi\nu}(r_2)u_{\pi}v_{\pi}\Bigl(
\displaystyle\frac{v^2_{\nu}}{\omega-E_{\pi\nu}}+\frac{u^2_{\nu}}{\omega+E_{\pi\nu}}\Bigr)\nonumber\\
&
\nonumber\\
&+&\sum\limits_{\pi<,(\nu)}t^2_{(\pi)(\nu)}\chi_{\pi}(r_1)\chi_{\pi}(r_2)u_{\pi}v_{\pi}g'_{(\nu)}(r_1,r_2,\mu_n-\omega-E_{\pi}),
\nonumber\\
&
\nonumber\\
A_{33}(r_1,r_2,\omega)&=&\sum\limits_{\pi<,\nu<}\chi_{\pi\nu}(r_1)\chi_{\pi\nu}(r_2)\Bigl(
\displaystyle\frac{u^2_{\pi}u^2_{\nu}}{\omega-E_{\pi\nu}}-\frac{v^2_{\pi}v^2_{\nu}}{\omega+E_{\pi\nu}}\Bigr)\nonumber\\
&
\nonumber\\
&+&\sum\limits_{\pi<,(\nu)}t^2_{(\pi)(\nu)}\chi_{\pi}(r_1)\chi_{\pi}(r_2)u^2_{\pi}g'_{(\nu)}(r_1,r_2,\omega-E_{\pi}+\mu_n)
\nonumber\\
&
\nonumber\\
&+&\sum\limits_{\nu<,(\pi)}t^2_{(\pi)(\nu)}\chi_{\nu}(r_1)\chi_{\nu}(r_2)u^2_{\nu}g'_{(\pi)}(r_1,r_2,\omega-E_{\nu}+\mu_p).
\label{propcont}
\eea
Here, $\pi<$ ($\nu<$) means $\pi\le\pi_{max}$ ($\nu\le\nu_{max}$), where $\pi_{max}$ 
($\nu_{max}$) is taken so that $E_{\pi_{max}}=\varepsilon_{\pi_{max}}-\mu_{p}$ 
($E_{\nu_{max}}=\varepsilon_{\nu_{max}}-\mu_{n}$) with an acceptable accuracy, and  the projected radial s-p Green function is:
\be
g'_{(\lambda)}(r_1,r_2,\varepsilon)=g_{(\lambda)}(r_1,r_2,\varepsilon)-
\sum\limits_{\lambda_<}\frac{\chi_\lambda(r_1)\chi_\lambda(r_2)}{\varepsilon-\varepsilon_\lambda}.
\label{greenf}
\ee
The first sum in this expression is running over
all the s-p states and can be calculated via a product of the regular and irregular solutions
of the corresponding s-p Schr\"odinger equation. 
Such a representation of $g_{(\lambda)}(r_1,r_2,\varepsilon)$ has been used in Ref.~\cite{10} to formulate the very first 
version of the continuum-RPA. 
Other $A_{KK'}(r,r',\omega)$ components in Eq. (\ref{propcont}) are obtained in the same way, as it is done in 
Eq. (\ref{propdiscr}). Thus, Eq. (\ref{propcont}) together with Eqs. (\ref{greenf}) and (\ref{eqexf}) realize a version 
of the pn-cQRPA method. As applied to description of the Fermi and GT strength distributions in single-open-shell nuclei, 
the method was used in Ref.~\cite{5}. This pn-cQRPA approach has been applied to description of the $\beta\beta$-decay matrix elements in Ref.~\cite{6}.

Within the pn-cQRPA the spectral decomposition of Eqs. (\ref{eqN8})-(\ref{eqN10}) remains valid, provided the continuum-states 
$|s\rangle$ (normalized to the $\delta$-function of the energy) are taken into consideration. These states are 
characterized, in particular, by the particle-hole (hole-particle) energy-dependent transition densities which determine 
the strength functions as follows:
\be
S^{(\mp)}_{JLS}(\omega)=(2J+1)\left | \int V_L(r) \varrho^{(\mp)}_{JLS}(r,\omega) dr \right |^2.
\label{sftrd}
\ee
These transition densities can be expressed in terms of the effective fields corresponding to the probing operators 
of Eq. (\ref{probop}) and their partners. Supposing the p-h
interaction is 
of zero range (with the intensities $F^{S}_{1}=F^{S}_{2}=2F^{S}$),
one gets from Eqs.~(\ref{eqexf})-(\ref{strf+}) the expressions:
\be
\varrho^{(-)}_{JLS}(r,\omega) = -\frac 1\pi \Image \frac{(2J+1)^{1/2}\tilde V^{(JLS)}_{1[1]}(r,\omega)}{2F^{S} \sqrt{S^{(-)}_{JLS}(\omega)}} \ ,
\varrho^{(+)}_{JLS}(r,\omega) = -\frac 1\pi \Image \frac{(2J+1)^{1/2}\tilde V^{(JLS)}_{2[2]}(r,\omega)}{2F^{S} \sqrt{S^{(+)}_{JLS}(\omega)}} \ .
\label{trd}
\ee
The intensities $F^{S}$ may be radial-dependent.

\section{Calculation of the GT strength functions and the $2\nu\beta\beta$-decay amplitudes within the pn-cQRPA}

\subsection{Model Hamiltonian: interactions in the particle-hole and particle-particle channels}

For quantitative realization of the relationships given in the preceding Section, we specify the nuclear Hamiltonian. 
We adopt here the realistic model Hamiltonian, which is
an extended version of that used in Refs.~\cite{5,6}. 
The Hamiltonian consists of the mean field and residual interactions. 
The partially selfconsistent mean field is described in the Subsect. B. 
Here, we specify interactions in the p-h 
and p-p channels, $\hat{H}_{p-h}$ and $\hat{H}_{p-p}$, respectively. 
As $\hat{H}_{p-h}$, we use the Landau-Migdal forces in the the charge-exchange channels:
\be
\hat{H}_{p-h}=2\sum_{a>b}(F^0+F^1\sss_a\sss_b)
\tau^{(-)}_a\tau^{(+)}_b\delta({\bf r}_a-{\bf r}_b)+h.c.,
\label{lmf}\ee
where the intensities $F^{S}$ of the non-spin-flip ($S=0$) and spin-flip 
($S=1$) parts of this interaction are the phenomenological parameters. 
We choose the zero-range p-p interaction in both the neutral (pairing) and 
charge-exchange channels:
\be
\hat{H}_{p-p}=-\frac{1}{2}\sum\limits_{\scriptsize \begin{array}{c}JLSM\\\lambda\lambda'\lambda_1\lambda_1'\end{array}}
G^{S}
(\lambda\lambda'\lambda_1\lambda_1')_{JLS}\Bigl(P^{JM}_{\lambda\lambda'}\Bigr)^\dagger P^{JM}_{\lambda_1\lambda_1'},
\label{ppf}\ee
where $G^{S}$ are the intensities of the p-p interaction in the $S=0,1$ channels, and (differently from the below-given
equations) the sums over s-p states include the neutron and proton states, 
$(\lambda\lambda'\lambda_1\lambda_1')_{JLS}=\int\chi^{JLS}_{\lambda,\lambda'}(r)\chi^{JLS}_{\lambda_1,\lambda_1'}(r) \frac{dr}{r^2}$.
The annihilation operator $P^{JM}$ for a nucleon pair is given, as follows:
\be
P^{JM}_{\lambda\lambda'}=\sum_{m,m'} \langle JM\vert j_{\lambda}m,j_{\lambda'}m'\rangle a_{\lambda'm'}a_{\lambda m},
\label{pairop}\ee
where $a_{\lambda m}$ ($a^{\dagger}_{\lambda m}$) is the annihilation (creation) operator of a nucleon in the state with the 
quantum numbers $\lambda m$ ($m$ is the projection of the particle angular momentum), 
$\langle JM\vert j_{\lambda}m,j_{\lambda'}m'\rangle$ is the Clebsh-Gordan coefficient. 
The p-h interaction of Eq. (\ref{lmf}) can also be expressed in terms of the operators $a^{\dagger}$ and $a$:
\be
\hat{H}_{p-h}=2\sum\limits_{\scriptsize \begin{array}{c}JLSM\\\pi\nu\pi'\nu'\end{array}}
F^{S}
(\pi\nu\pi'\nu')_{JLS}
\Bigl(A^{JM}_{\pi\nu}\Bigr)^\dagger A^{JM}_{\pi'\nu'}
+\biggl(\pi\leftrightarrow\nu\biggr),
\label{phint}\ee
where the creation p-h operator $A^{\dagger}$ is given as follows:  
\be
\Bigl(A^{JM}_{\pi\nu}\Bigr)^\dagger=\sum_{m_{\pi},m_{\nu}}
\langle JM\vert j_{\pi}m_{\pi},j_{\nu}-m_{\nu}\rangle(-1)^{j_{\nu}-m_{\nu}}a^\dagger_{\pi m_{\pi}}a_{\nu m_{\nu}}.
\label{phop}\ee
In solution of the pairing problem we simplify the p-p interaction (\ref{ppf}) for
the $0^{+}$ neutral channel, using the ``diagonal" approximation: $\lambda=\lambda'$,
$\lambda_{1}=\lambda'_{1}$. The nucleon pairing is described with the use of the 
Bogolyubov transformation in terms of the quasiparticle creation (annihilation) operators
$\alpha_{\lambda m}^{\dagger}$ , $\alpha_{\lambda m}$
(see, e.g., Ref. \cite{11}). 
As a result, the following model Hamiltonian is obtained to describe charge-exchange 
excitations within the pn-QRPA:
\be
\hat{H}=\hat{H}_{0,n}+\hat{H}_{0,p}+\mu_n\hat{N}+\mu_p\hat{Z}+\hat{H}_{p-h}+\hat{H}_{p-p},\ \ \ \ \ 
\hat{H}_0=\sum_{\lambda m}E_{\lambda}\alpha_{\lambda m}^\dagger \alpha_{\lambda m}.
\label{qpham}\ee
Here, $E_{\lambda}=\sqrt{ (\varepsilon_{\lambda}-\mu)^2+
\Delta_{\lambda}^2}$ is the quasiparticle energy. The chemical potentials
$\mu$ and the energy-gap parameters $\Delta_{\lambda}$ are determined 
by solving the BCS-model equations for even and odd subsystems (with the number of nucleons 
$N(even)$ and $N(odd)$ nucleons, respectively):
\be
N(even)=\sum_{\lambda}(2j_{\lambda}+1)v^2_{\lambda},\ \ \ \Delta_{\lambda}(even)=
G^0\sum_{\lambda'}(2j_{\lambda'}+1)(\lambda\lambda\lambda'\lambda')_{000}\frac{\Delta_{\lambda'}}{2E_{\lambda'}},
\label{Neven}
\ee
\be
N(odd)=\sum_{\lambda\ne\lambda_0}(2j_{\lambda}+1)v^2_{\lambda}+(2j_{\lambda_0}-1)v^2_{\lambda_0}+1,
\label{Nodd}
\ee
\be
\Delta_{\lambda}(odd)=
G^0\biggl\{\sum_{\lambda'\ne\lambda_0}(2j_{\lambda'}+1)(\lambda\lambda\lambda'\lambda')_{000} \frac{\Delta_{\lambda'}}{2E_{\lambda'}}
+(2j_{\lambda_0}-1)(\lambda\lambda\lambda_0\lambda_0)_{000}\frac{\Delta_{\lambda_0}}{2E_{\lambda_0}}
\biggr\}.
\label{bcseqs}\ee
Here, $\lambda_0$ labels the level occupied by the odd qausiparticle. Following the standard procedure, we introduce different strengths $G^{0}_{n}$ and $G^{0}_{p}$ of the p-p interaction in the $0^{+}$ pairing channels to reproduce the experimental pairing energies. 
The latter are expressed via the ground-state binding energies calculated within the BCS model:
\be
-{\cal E}_b(even)= \sum_{\lambda}(2j_{\lambda}+1)v^2_{\lambda}(\mu-E_{\lambda}),
\label{E0even}\ee
\be
-{\cal E}_b(odd)=\sum_{\lambda\ne\lambda_0}(2j_{\lambda}+1)v^2_{\lambda}(\mu-E_{\lambda})+
(2j_{\lambda_0}-1)v^2_{\lambda}(\mu-E_{\lambda_0})+{\varepsilon}_{\lambda_0}.
\label{bcsgsen}\ee
The level $\lambda_0$ is chosen from the minimum of the calculated $-{\cal E}_b$, and/or accordingly to the quantum numbers of the odd-nucleus ground state. To roughly correct for double counting in evaluation of the total energy, we properly 
modify Eqs. (\ref{E0even}),(\ref{bcsgsen}): $(\mu-E_{\lambda})\to(\mu-E_{\lambda}-\frac{1}{2}U_{\lambda\lambda})$; 
$\varepsilon_{\lambda_0}\to\varepsilon_{\lambda_0}-\frac{1}{2}U_{\lambda_0\lambda_0}$. Here $U_{\lambda\lambda}$ is 
the diagonal matrix element of the nuclear mean field (Subsect. B.).

The explicit expressions for the p-h and p-p interactions in terms of the quasiparticle (pn)-pair creation and 
annihilation operators are:
\be
\hat{H}_{p-h}=2\sum\limits_{\scriptsize \begin{array}{c}JLSM\\\pi\nu\pi'\nu'\end{array}}
F^{S}
(\pi\nu\pi'\nu')_{JLS}
\Bigl(Q^{JM}_{\pi\nu}\Bigr)^\dagger Q^{JM}_{\pi'\nu'}+\biggl(p\leftrightarrow n, \pi\leftrightarrow\nu\biggr),
\label{hphqp}\ee
\be
\hat{H}_{p-p}=-\sum\limits_{\scriptsize \begin{array}{c}JLSM\\\pi\nu\pi'\nu'\end{array}}
G^{S}
(\pi\nu\pi'\nu')_{JLS}
\Bigl({\cal P}^{JM}_{\pi\nu}\Bigr)^\dagger{\cal P}^{JM}_{\pi'\nu'},
\label{hppqp}\ee
where
\be
Q^{JM}_{\pi\nu}=u_{\pi}v_{\nu}{\cal A}^{JM}_{\pi\nu}+
v_{\pi}u_{\nu}\Bigl({\cal A}^{J\widetilde{M}}_{\pi\nu}\Bigr)^\dagger,\ \ \ 
{\cal P}^{JM}_{\pi\nu}=u_{\pi}u_{\nu}{\cal A}^{JM}_{\pi\nu}-
v_{\pi}v_{\nu}\Bigl({\cal A}^{J\widetilde{M}}_{\pi\nu}\Bigr)^\dagger,
\label{qppairop1}\ee
\be
{\cal A}^{JM}_{\pi\nu}=\sum_{mm'}\langle JM\vert j_{\pi}m,j_{\nu}m'\rangle\alpha_{\nu m'}\alpha_{\pi m},
\ \ \ {\cal A}^{J\widetilde{M}}_{\pi\nu}=(-1)^{J-M}{\cal A}^{J-M}_{\pi\nu}.
\label{qppairop2}\ee
The forward and backward amplitudes $X$ and $Y$ are defined by the corresponding matrix elements 
$X^{JLS}_{s,\pi\nu}=\langle s,J^\pi|({\cal A}^{JM}_{\pi\nu})^{+}|0 \rangle$ and
$Y^{JLS}_{s,\pi\nu}=\langle s,J^\pi|{\cal A}^{J\tilde M}_{\pi\nu}|0 \rangle$. The system of equations for 
these amplitudes can be explicitly derived using the model Hamiltonian of 
Eqs. (\ref{qpham}), (\ref{hphqp})-(\ref{qppairop2}). Having transformed this system to the
form given by Eqs. (\ref{defvarrho}),(\ref{rhocoord}),(\ref{propdiscr}), one can relate the interaction 
$F^{S}_{K}(r,r')$, entering the basic pn-QRPA Eqs. (\ref{rhocoord}),
(\ref{eqexf}), to the model-Hamiltonian zero-range interactions of Eqs. (\ref{hphqp})-(\ref{qppairop2}):
\be
F^S_K(r,r')=F^S_K\,\frac{\delta(r-r')}{rr'}, \ \ \ F^S_{K=1,2}=2F^S,
\ \ \ F^S_{K=3,4}=-G^S.
\label{intstr}\ee
Thus, within the model under consideration there are four phenomenological interaction
strength parameters $F^S$, $G^S$ ($S=0,1$) which will be specified below.

\subsection{Model Hamiltonian: the mean field and selconsistency conditions}

The nuclear mean field consists of the isoscalar (central and spin-orbit), isovector
and Coulomb parts:
\be
U(x)=U_0(x)+U_1(x)+U_C(x),\ \ \ U_0(x)=U_0(r)+U_{SO}(r){\bf l}\sss,
\label{meanff}\ee
\be
U_1(x)=\frac{1}{2}v(r)\tau^{(3)},\ \ \ U_C(x)=\frac{1}{2}(1-\tau^{(3)})U_C(r),
\label{meanf}\ee
where $v(r)$ is the symmetry potential. Within the model the mean-field isoscalar part
is treated  phenomenologically (that seems 
well justified for description of isovector excitations), while the isovector and Coulomb parts are calculated selfconsistently.

The realistic phenomenological isoscalar mean field is parametrized as follows:
\be
U_0(r)=-U_0f_{WS}(r,R,a),\ \ \ U_{SO}(r)=U_{SO}\frac{1}{r}\frac{df_{WS}}{dr},
\label{isosmf}\ee
where $f_{WS}=(1+exp(\frac{r-R}{a}))^{-1}$ is the Woods-Saxon function with the radius
$R=r_0A^{1/3}$ and the diffuseness parameter $a$. To avoid spurious violation of the isospin
symmetry caused by the mean-field isovector part, the latter is calculated using the
isospin-selfconsistency condition~\cite{5}. This condition relates the symmetry potential to
the isovector nuclear density (the neutron-excess density) via the isovector (non-spin-flip) 
part of the p-h interaction (\ref{lmf}):
\be
v(r)=2F^{0}(n^n(r)-n^p(r)).
\label{isovscc}\ee
Here, the neutron and proton densities are calculated with taking into account nucleon pairing:
\be
n_{even}(r)=\frac{1}{4\pi r^2}\sum_{\lambda}(2j_{\lambda}+1)
v^2_{\lambda}\chi^2_{\lambda}(r),
\label{nucldeneven}\ee
\be
n_{odd}(r)=\frac{1}{4\pi r^2}\biggl\{\sum_{\lambda\ne\lambda_0}(2j_{\lambda}+1)
v^2_{\lambda}\chi^2_{\lambda}(r)+
\Bigl[(2j_{\lambda_0}-1)v^2_{\lambda_0})+1\Bigr]\chi^2_{\lambda_0}(r)\biggr\}.
\label{nucldenodd}\ee
The mean Coulomb field $U_{C}(r)$ is also calculated selconsistently within the Hartree approximation via the proton 
density (\ref{nucldeneven}),(\ref{nucldenodd}). Thus, the above-described mean field is a partially selfconsistent one. Nevertheless, the 
isospin symmetry of the model Hamiltonian is slightly violated due to approximate description of the nucleon pairing 
(the use of the ``diagonal" approximation, truncation of the s-p basis, different values of the $G^0_{n}$ and 
$G^0_{p}$ pairing strengths). These approximations can affect the quality of the calculated Fermi excitations, but they 
seem to be of rather minor importance for calculations of the GT excitations considered in this work.

\subsection{Choice of model parameters and the calculation scheme}

Recently, a partially selfconsistent mean field has been adopted to describe experimental single-particle spectra in the 
doubly-closed-shell nuclei $^{48}$Ca, $^{132}$Sn, $^{208}$Pb \cite{7}. 
As a starting point, the fully phenomenological mean field of Ref.~\cite{12} was used. 
The deduced strength parameters $U_0$, $U_{SO}$, $F^0=Cf'$ ($C=300\ MeV\cdot fm^3$) 
and also the diffuseness parameter $a$ reveal a weak A-dependence (the value $r_{0}$=1.27 fm is taken the same for all nuclei \cite{12}). This point allows us to use for the nuclei under consideration the properly interpolated values 
of the model parameters mentioned above (Table 1).
Then, the partially selfconsistent mean field (\ref{meanff})-(\ref{nucldenodd}) and the corresponding s-p level scheme are 
calculated to get a basis for the BCS solution of the nucleon pairing problem accordingly to Eqs. (\ref{Neven})-(\ref{bcseqs}). This problem is solved consistently
with calculation of the symmetry potential and mean Coulomb field by making use of the neutron and proton densities (\ref{nucldeneven}), (\ref{nucldenodd}) and a set of the pairing strength parameters $G^0_{\alpha}$. The latter are then fitted to reproduce the experimental pairing energies ${\cal E}_{pair}$~\cite{13}:
\be
{\cal E}_{pair}(N)=\pm\biggl\{\frac{1}{8}{\cal E}_b(N+2)-\frac{1}{2}{\cal E}_b(N+1)+
\frac{3}{4}{\cal E}_b(N)-\frac{1}{2}{\cal E}_b(N-1)+\frac{1}{8}{\cal E}_b(N-2)\biggr\}
\label{pairen}\ee
with ``+" or ``-" related to $N(even)$ or $N(odd)$, respectively. As ${\cal E}_b$ in this expression (the corresponding experimental values are taken from Ref.~\cite{14}), we use the properly modified ground-state energies of Eqs.~(\ref{E0even}),(\ref{bcsgsen}). For the subsystems having $N_{c}\pm 2$ nucleons ($N_c$ is a magic number) we use a simplified expression for the pairing energy:
\be
{\cal E}_{pair}(N_{c}\pm 2)=-\frac{1}{4}{\cal E}_b(N_{c}\pm 1)+\frac{3}{4}{\cal E}_b(N_{c}\pm 2)-
\frac{3}{4}{\cal E}_b(N_{c}\pm 3)+\frac{1}{4}{\cal E}_b(N_{c}\pm 4) .
\label{pairenc}\ee

The parameters $G^0_{n,p}=Cg^0_{n,p}$, fitted as described above, are listed in Table 1. 
To reduce an artificial violation of the isospin symmetry, caused by the use of only discrete (quasidiscrete)
s-p states in description of the pairing problem, we use the same number of these states
$N_{d+qd}$ for the proton and neutron subsystems. The $N_{d+qd}$ values for the nuclei under consideration are 
also given in Table 1. The energy and the radial wave function
of a quasibound state $\lambda$ are found using the representation of the radial Green
function, which is valid inside the nucleus at the energies close to the quasibound state energy:
\be
g_{(\lambda)}(r,r',\varepsilon)=
\frac{\chi_{\lambda}(r)\chi_{\lambda}(r')}{\varepsilon-\varepsilon_{\lambda}+\frac{{\rm i}}{2}\Gamma_{\lambda}},
\label{gfatqbse}\ee
where $\Gamma_{\lambda}$ is the quasibound-state escape width.

After the consistent solution of the pairing problem, we turn to evaluation of the
GT$^{\pm}$ strength functions and amplitudes $M^{2\nu}_{GT}$ within the present
pn-cQRPA approach. To calculate these quantities in accordance with Eqs. (\ref{eqexf}), (\ref{strf-}), 
(\ref{strf+})
and Eqs. (\ref{ndstrf}), (\ref{17}), (\ref{19}), respectively one has to choose the strengths
$F^1=Cg'$ and $G^1$ (or $g_{pp}=2G^1/(G^0_n+G^0_p)$) of the spin-flip parts of the
p-h and p-p interactions. 
Using in evaluation of the GT$^{-}$ strength function the realistic value $g_{pp}=1$, we find
the strength $g'$ to reproduce in calculations the experimental GTR energy. 
The fitted strength values are listed in Table 2. 
These values are weakly dependent on the variation of $g_{pp}$ in a vicinity of the point $g_{pp}=1$.
Having chosen $g'$, the amplitudes $M^{2\nu}_{GT}$, their
components $M'_{GT}$ and $M''_{GT}$ are calculated as functions of $g_{pp}$ for the pairs of nuclei $^{76}$Ge-Se, 
$^{100}$Mo-Ru, $^{116}$Cd-Sn and $^{130}$Te-Xe. 
Since the $^{116}$Sn nucleus has the filled proton shell, in evaluation of the amplitude $M^{2\nu}_{GT}$ 
for the pair $^{116}$Cd-Sn we identify the vacuum $|0'\rangle$ with that for $|0\rangle$.
The strengths $g_{pp}\sim 1$, which allow us to reproduce in calculations the experimental $M^{2\nu}_{GT}$
values, are also listed in Table 2. These strengths are further used to 
evaluate the $2\nu\beta\beta$-decay GT reduced strength function 
$S^{2\nu}_{GT}(\omega)=\omega^{-1}S^{(-\,-)}_{GT}(\omega)/M^{2\nu}_{GT}$ 
(normalized to unity) and the corresponding reduced running sum 
$r^{2\nu}_{GT}(\omega)=\int^{\omega}S^{2\nu}_{GT}(\omega)d\omega$ 
(goes to unity at large $\omega$). Along with the above-mentioned decomposition
these quantities show how the amplitudes $M^{2\nu}_{GT}$ are formed from the partial 
contribution of $1^+$ states in the intermediate nucleus. 
Finally, the fitted $g'$ and $g_{pp}$ values are used to calculate the GT$^{\mp}$ strength functions. 
To compare with the experimental GT$^{-}$ strength distribution it is convenient to consider the 
reduced running sum $r^{(-)}_{GT}(\omega)=\int^{\omega} S^{(-)}_{GT}(\omega)d\omega /S^{(-)}_{GTR}$,
where $S^{(-)}_{GTR}$ is the GT$^{-}$ strength of the GTR.
We show also the (non-reduced) running sum $R^{(+)}_{GT}(\omega)=\int^{\omega}S^{(+)}_{GT}(\omega)d\omega$
for the GT$^{+}$ strength in the final nucleus. 
All the calculations are performed without taking  into account the quenching of the total 
GT strength.

\section{Calculation results}

Within the pn-cQRPA approach described in the preceding Sections we calculate the $2\nu\beta\beta$-decay amplitudes 
$M^{2\nu}_{GT}$ and the corresponding GT$^{\pm}$ strength functions for the pairs of nuclei $^{76}$Ge-Se, 
$^{100}$Mo-Ru, $^{116}$Cd-Sn and $^{130}$Te-Xe.
The results are presented in Figs. 1.-4. 
Each figure contains: 
a) the amplitude $M^{2\nu}_{GT}$ and its decomposition in two terms in accordance with 
Eq. (\ref{19}) calculated as functions of the strength $g_{pp}$; 
b) the reduced $2\nu\beta\beta$-decay GT strength function $S^{2\nu}_{GT}(\omega)$ calculated at the $g_{pp}$ 
value fitted to reproduce in calculations the experimental amplitude $M^{2\nu}_{GT}$ (Table 2.); 
c) the reduced $2\nu\beta\beta$-decay running sum $r^{2\nu}_{GT}(\omega)$ calculated at the fitted $g_{pp}$ value; 
d) the calculated GT$^{-}$ strength function $S^{(-)}_{GT}(\omega)$ for the initial nucleus, 
e) and f) the running sums $r^{(-)}_{GT}$ and $R^{(+)}_{GT}$ for the initial and final nuclei, respectively.
The results are compared with available experimental data. 

The calculated results shown in Figs. 1.a), b), c) - 4.a), b), c) demonstrate how the amplitudes $M^{2\nu}_{GT}$ are 
formed for the nuclei under consideration. 
The general features are the almost linear dependence of the component $M''_{GT}$ on the p-p interaction strength 
$g_{pp}$ and vanishing of these amplitudes at $g_{pp}\approx 1$. 
Thus, we confirm the conclusion of Ref.~\cite{41} (obtained within a simpler model) that the component 
$M''_{GT}$ is only determined by the p-p residual interaction and goes to zero at $g_{pp}=1$ where the spin-isospin 
SU(4)-symmetry is restored in the p-p sector of this model Hamiltonian \footnote{In the QRPA calculations using a realistic G-matrix residual interaction~\cite{19} this point of crossing zero is shifted to $g_{pp}<1$ due to the effect of SU(4)-symmetry breaking by the tensor forces in the residual interaction}. 
As compared with Ref.~\cite{41}, in our calculations the components $M'_{GT}$ (and, therefore, the amplitudes 
$M^{2\nu}_{GT}$) are somewhat smaller since the strength of the mean-field spin-orbit term used in our calculations is smaller. 
Again we confirm the conclusion of Ref.~\cite{41} that at $g_{pp}\sim 1$ the component $M'_{GT}$ essentially 
depends on the spin-orbit term only, which is the main source of SU(4)-symmetry breaking in nuclei.
Details of the formation of the amplitudes $M^{2\nu}_{GT}$ are dependent on the shell-structure of nuclei. 
As follows from Figs.
2.b), 2.c) and 3.b), 3.c), population of the 1$^{+}$ ground state of the intermediate nuclei $^{100}$Tc and $^{116}$In 
dominates the formation of the amplitudes $M^{2\nu}_{GT}$ for the pairs $^{100}$Mo-Ru and $^{116}$Cd-Sn, respectively. 
Within the used calculation scheme the proper GT$^{-}$ and GT$^{+}$ transitions are mainly to the single-quasiparticle 
g$^{n}_{9/2}\ \to$ g$^{p}_{7/2}$ and g$^{p}_{7/2}\ \to$ g$^{n}_{9/2}$ transitions, respectively. 
For the pairs $^{76}$Ge-Se and $^{130}$Te-Xe the amplitudes $M^{2\nu}_{GT}$ are formed by population of many 1$^{+}$ 
states of the intermediate nuclei $^{76}$As and $^{130}$I, respectively (Figs. 1.b), 1.c) and 4.b), 4.c)). 
It means, in particular, that the intermediate states having a relatively large excitation energy and, therefore, 
a small B(GT$^{+}$) value (like the GTR) can, nevertheless, 
play essential role in the formation of the amplitude $M^{2\nu}_{GT}$. 
For this reason, the experimental studies aimed to deduce the corresponding B(GT$^{\pm}$) values for the 
intermediate 1$^{+}$ states and then to reconstruct the amplitudes $M^{2\nu}_{GT}$ seem to be sufficient 
in the case of the approximate single-state dominance.
The proper examples are the studies performed for 1$^{+}$ states in $^{100}$Tc ~\cite{16} and $^{116}$In ~\cite{3}, 
but it might not be fully correct in the case of the study of $^{76}$As ~\cite {4}. 

Turning to calculations of the GT$^{(\mp)}$ strength functions, we start from the description of the GTR in the intermediate
nuclei $^{76}$As, $^{100}$Tc, $^{116}$Sb, $^{130}$I. 
The strength functions $S^{(-)}_{GT}(\omega)$ are shown in Figs. 1.d)-4.d) (similarly to the strength functions shown in 
Figs. 1.b)-4.b) we add a small imaginary part to the s-p potential while calculating the s-p Green's function in order
to make the presentation more visible). 
Using the experimental data of Refs.~\cite {15,16,17} on charge-exchange reactions, we fit the p-h strength parameter 
$g'$ to reproduce in calculations the experimental value of the centroid of the GTR energy (Table 2). 
For A=116 systems we have performed the fitting of the GTR in ${^{116}}$Sb in view of the absence of appropriate experimental data for $^{116}$In. 
However, the value B(GT$^{(-)}$)=$0.26\pm 0.02$ has been deduced in Ref.~\cite{2} from the $^{116}$Cd(p,n)-reaction with 
population of the 1$^+$ ground state in $^{116}$In.
Within the interval $E_x\le 3$ MeV the calculated GT$^{(-)}$-strength distribution in $^{116}$In exhibits one 1$^+$ state, 
corresponding to the back-spin-flip transition $1g^n_{7/2}\to 1g^p_{9/2}$ into the 1$^+$ ground state of $^{116}$In, with the value B(GT$^{(-)}$)=1.05 (for $^{116}$Sn$\to ^{116}$Sb this transition is Pauli blocked).
Bearing in mind that the low-energy part of the GT$^{(\mp)}$ strength distributions cannot be described in details 
within any pn-QRPA-based approach (for instance, in view of existence of the quasiparticle-phonon coupling), 
we compare with experimental data the running sums $r^{(-)}_{GT}(\omega)$, $R^{(+)}_{GT}(\omega)$. 
The calculated and experimental running sums $r^{(-)}_{GT}$ shown in Figs. 1.d)-4.d),
at least qualitatively, agree.
With the  exception for the running sums $R^{(+)}_{GT}$ for the pairs $^{76}$Se-$^{76}$As (Fig. 1.f)) and $^{116}$Sn-$^{116}$In (Fig. 3.f)),
a reasonable description of the data is found for the $B(GT^+)$ values corresponding to the ground-state to ground-state 
$1^+$ transitions $^{100}$Tc-$^{100}$Ru (Fig. 2.f) and $^{116}$Sn-$^{116}$In (Fig. 3.f).

\section{Conclusions}

We have given a detailed formulation of a version of the proton-neutron continuum-QRPA approach that makes use of:\\
(i) realistic zero-range interactions in the particle-hole and particle-particle channels; \\
(ii) a modern version of the phenomenological isospin-self-consistent nuclear mean field;\\
(iii) the full basis of the particle-hole excitations, and a rather large basis of the particle-particle ones (with 
inclusion of several quasi-bound s-p states).
Within the approach we have studied ``anatomy" of the Gamow-Teller $2\nu\beta\beta$-decay amplitude for a number 
of nuclei. The concept of the broken spin-isospin SU(4)-symmetry in nuclei has been made use of in the study of the $g_{pp}$-sensitivity of $M^{2\nu}_{GT}$.
A reasonable description of the Gamow-Teller strength functions in $\beta^{\mp}$-channels for the nuclei in question 
has been obtained within the approach.

Sensitivity of the calculated observables to realistic variations of the mean field parameters
is going to be studied elsewhere.

\acknowledgments

This work of S.Yu.I. and M.H.U. is supported in part by the Russian Foundation for Basic Researches under Contract No. 09-02-00926-a. V.A.R. and A.F. acknowledge support of the Deutsche Forschungsgemeinschaft within the SFB TR27 "Neutrinos and Beyond".

\begin{table}
\caption{The phenomenological mean field parameters, singlet p-h and p-p interaction 
strengths, and the number of the bound and quasibound s-p states used in calculations.}
\begin{tabular}{|l|c|c|c|c|c|c|c|c|c|c|}
    \hline  
Nuclei  & $U_0,$ MeV & $U_{SO},$ MeV$\cdot$fm$^2$ & $a$, fm  & $f'$ & $g_{0,n}$ & $g_{0,p}$ & $N_{b+qb}$ \\ \hline
\ $^{76}$Ge-$^{76}$Se   & 51.3      &  34.4   & 0.60 & 1.17 & 0.41     &  0.32    &    16        \\
$^{100}$Mo-$^{100}$Ru & 51.5      &  34.2   & 0.61 & 1.13 & 0.45     &  0.41    &    16        \\
$^{116}$Cd-$^{116}$Sn & 51.6      &  34.1   & 0.62 & 1.06 & 0.39     &  0.33    &    22        \\
$^{130}$Te-$^{130}$Xe & 51.7      &  34.0   & 0.63 & 1.09 & 0.36     &  0.36    &    22        \\ \hline
\end{tabular}
\end{table}

\

\begin{table}
\caption{The phenomenological triplet p-h and p-p interaction strengths. 
The experimental GTR energies from Refs. ~\cite{15}-\cite{17}
and the amplitudes $M^{2\nu}_{GT}$ from Ref. ~\cite{19Barabash} are also given.}
\begin{tabular}{|l|c|c|c|c|c|c|c|c|c|c|}   
 \hline  
Nuclei  & $E_{x,GTR}$, MeV & $g'$ & $M^{2\nu}_{GT}$, MeV$^{-1}$ & $g_{pp}$ \\ \hline
\ $^{76}$Ge-$^{76}$Se   & 11.13 ~\cite {15}   & 0.81  & 0.14    &  0.62  \\
$^{100}$Mo-$^{100}$Ru & 13.3  ~\cite {16}   & 0.92  & 0.24    &  0.91   \\
$^{116}$Cd-$^{116}$Sn & 10.04 ~\cite {17}   & 0.77   & 0.13    &  1.07  \\
$^{130}$Te-$^{130}$Xe & 13.59 ~\cite {15}   & 0.88   & 0.03    &  0.99  \\ \hline
\end{tabular}
\end{table}


\begin{figure}[htb]
\includegraphics[
width=8cm,height=14cm]{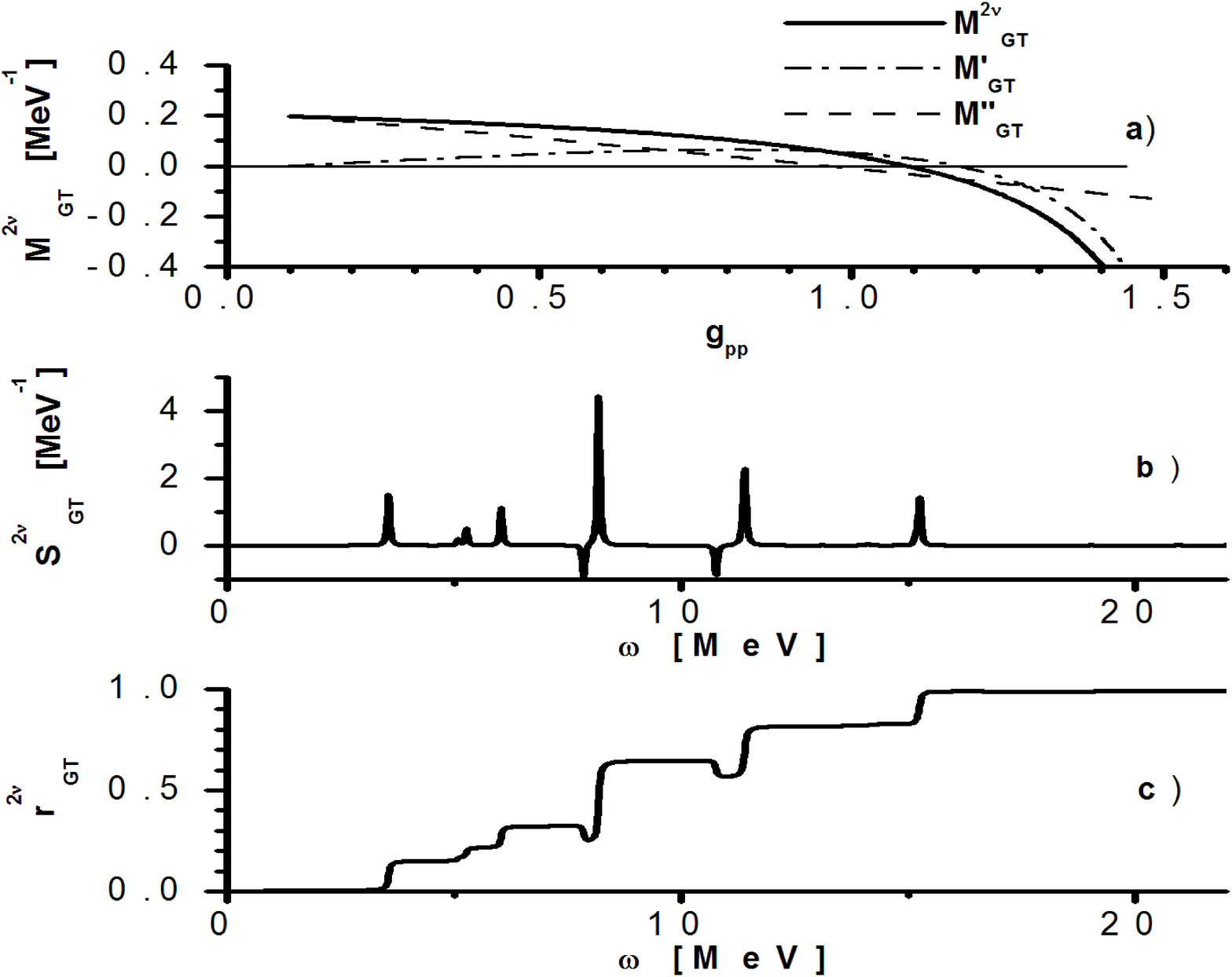}~~~~~\includegraphics[
width=8cm,height=14cm]{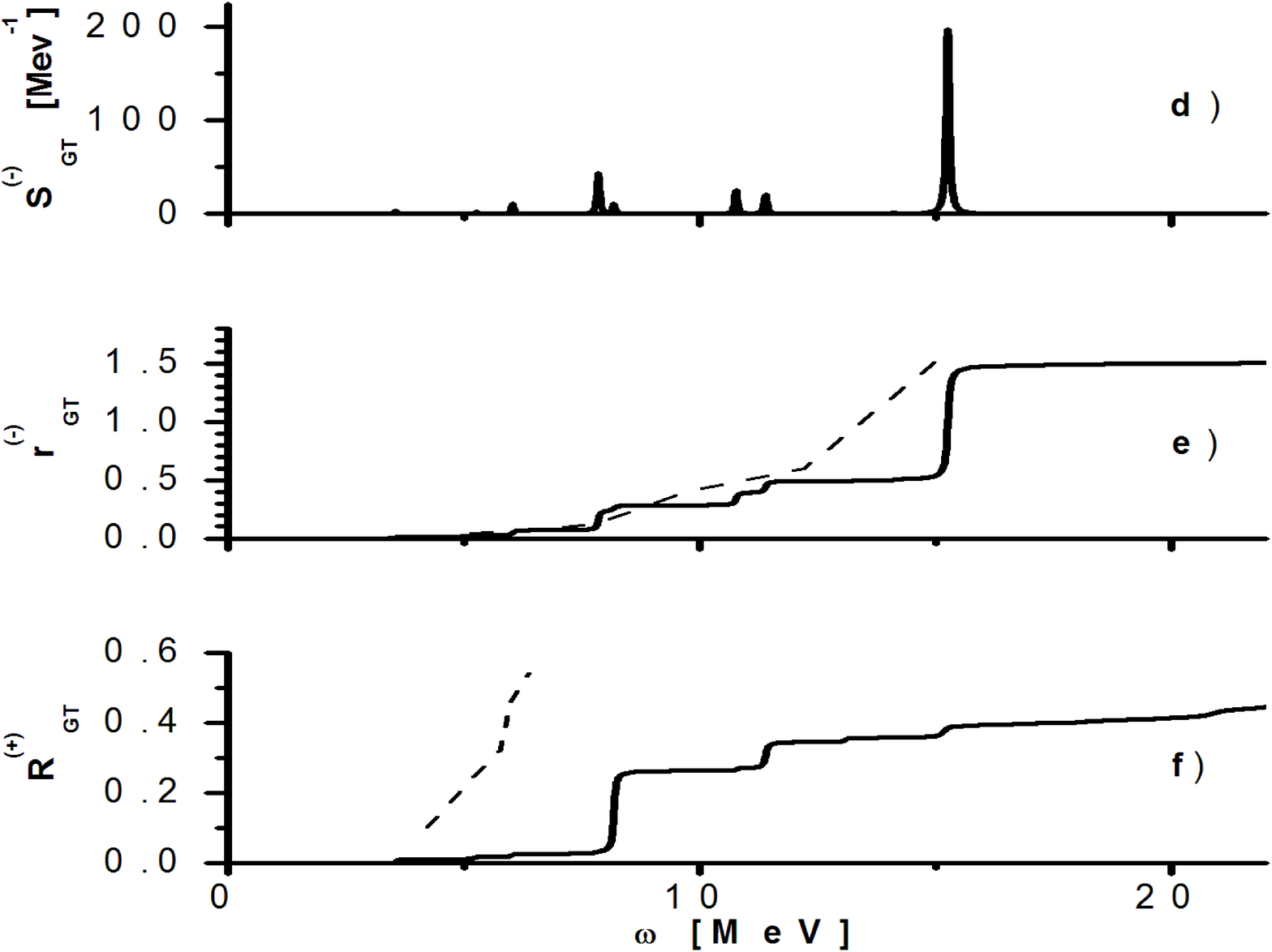}
\caption{ 
a) the $2\nu\beta\beta$-decay GT amplitude $M^{2\nu}_{GT}$ (full line) and their components $M'_{GT}$ 
(dotted-point line), $M''_{GT}$ (dotted line) calculated as a function of the strength $g_{pp}$ for the pair $^{76}$Ge-Se; 
b) and c) the $2\nu\beta\beta$-decay reduced GT strength function and running sum, respectively, calculated at the 
strength $g_{pp}$ from Table 2 for the same pair of nuclei; 
d) and e) the GT$^-$ strength function and reduced running sum, respectively, calculated at the strengths $g'$ and $g_{pp}$ 
from Table 2 for the pair $^{76}$Ge-As (full line), the experimental data (dashed line) are taken from Ref.~\cite{15}; 
f) the GT$^+$ running sum calculated at the strengths $g'$ and $g_{pp}$ 
from Table 2 for the pair $^{76}$Se-As (full line), the experimental data (dashed line) are taken from Ref.~\cite{4}.
}
\label{fig1}
\end{figure}

\begin{figure}[htb]
\includegraphics[
width=8cm,height=14cm]{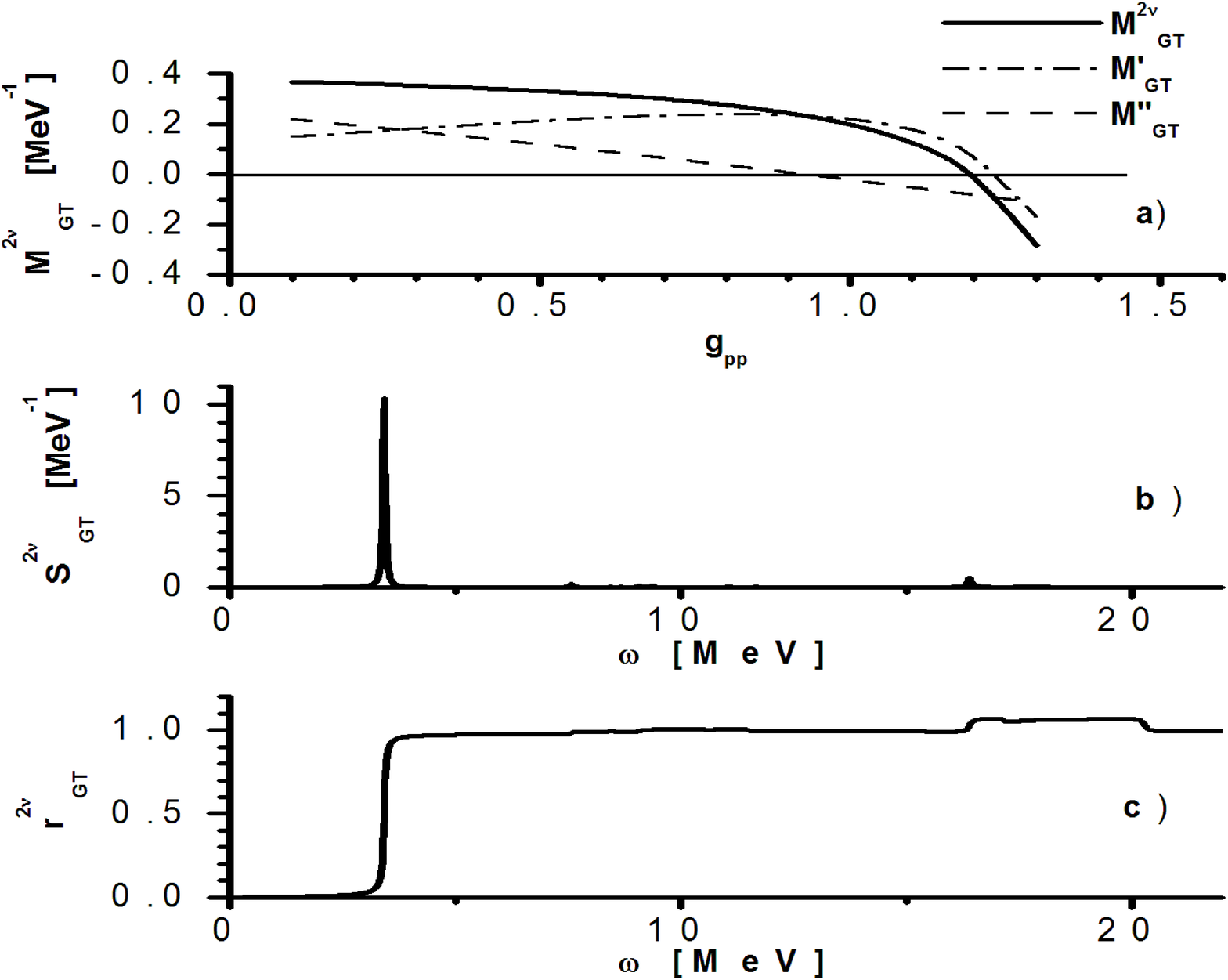}~~~~~\includegraphics[
width=8cm,height=14cm]{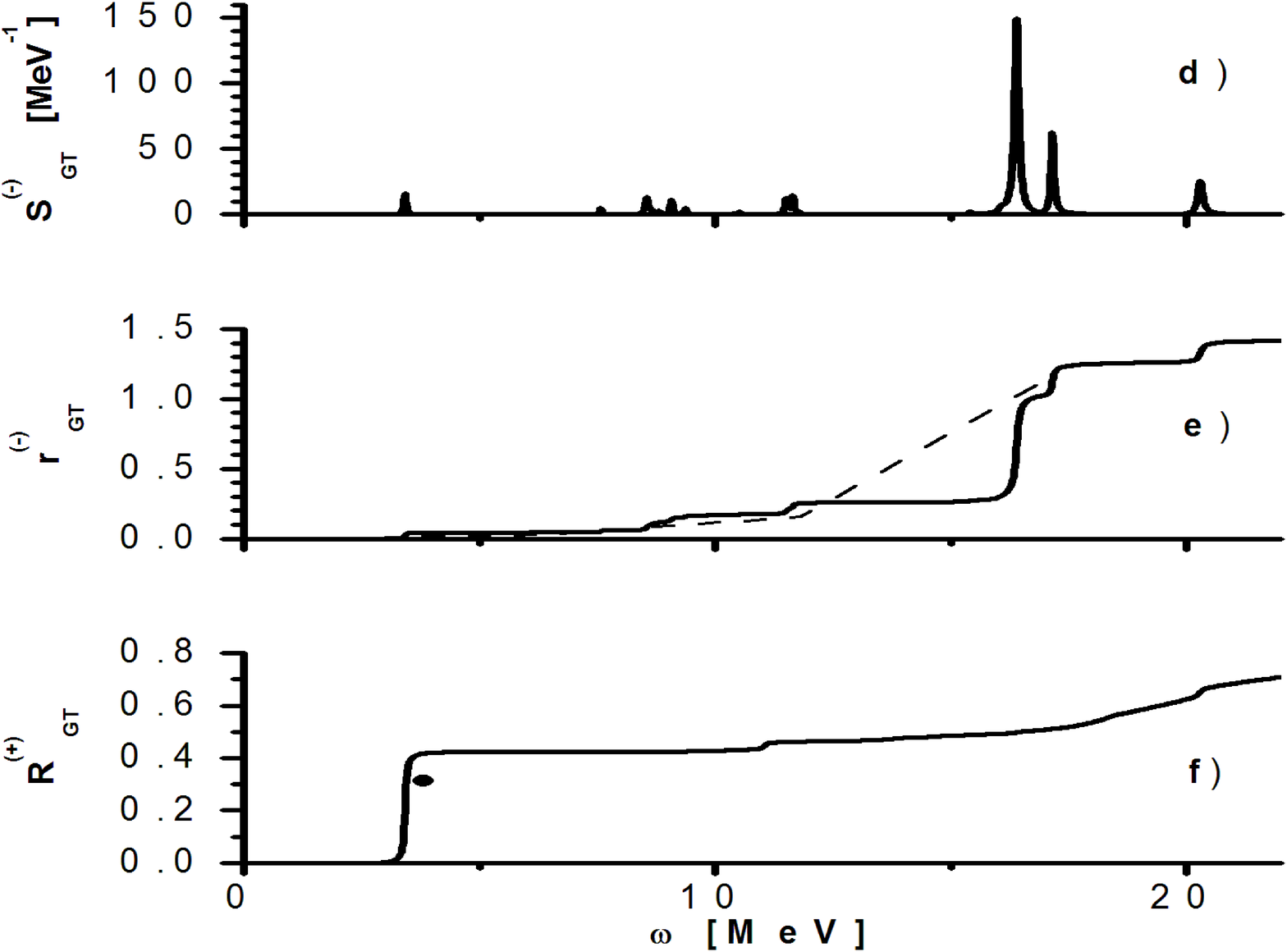}
\caption{ 
a)-c) the same as in Figs. 1.a)-1.c) but for the pair $^{100}$Mo-Ru; 
d) and e) the same as in Figs. 1.d) and 1.e) but for the pair $^{100}$Mo-Tc, 
the experimental data are taken from Ref.~\cite{16}; 
f) the same as in Fig. 1.f) but for the pair $^{100}$Ru-Tc, the experimental data (noted by the circle) 
are taken from Ref.~\cite{16}.
}
\label{fig2}
\end{figure}

\begin{figure}[htb]
\includegraphics[
width=8cm,height=14cm]{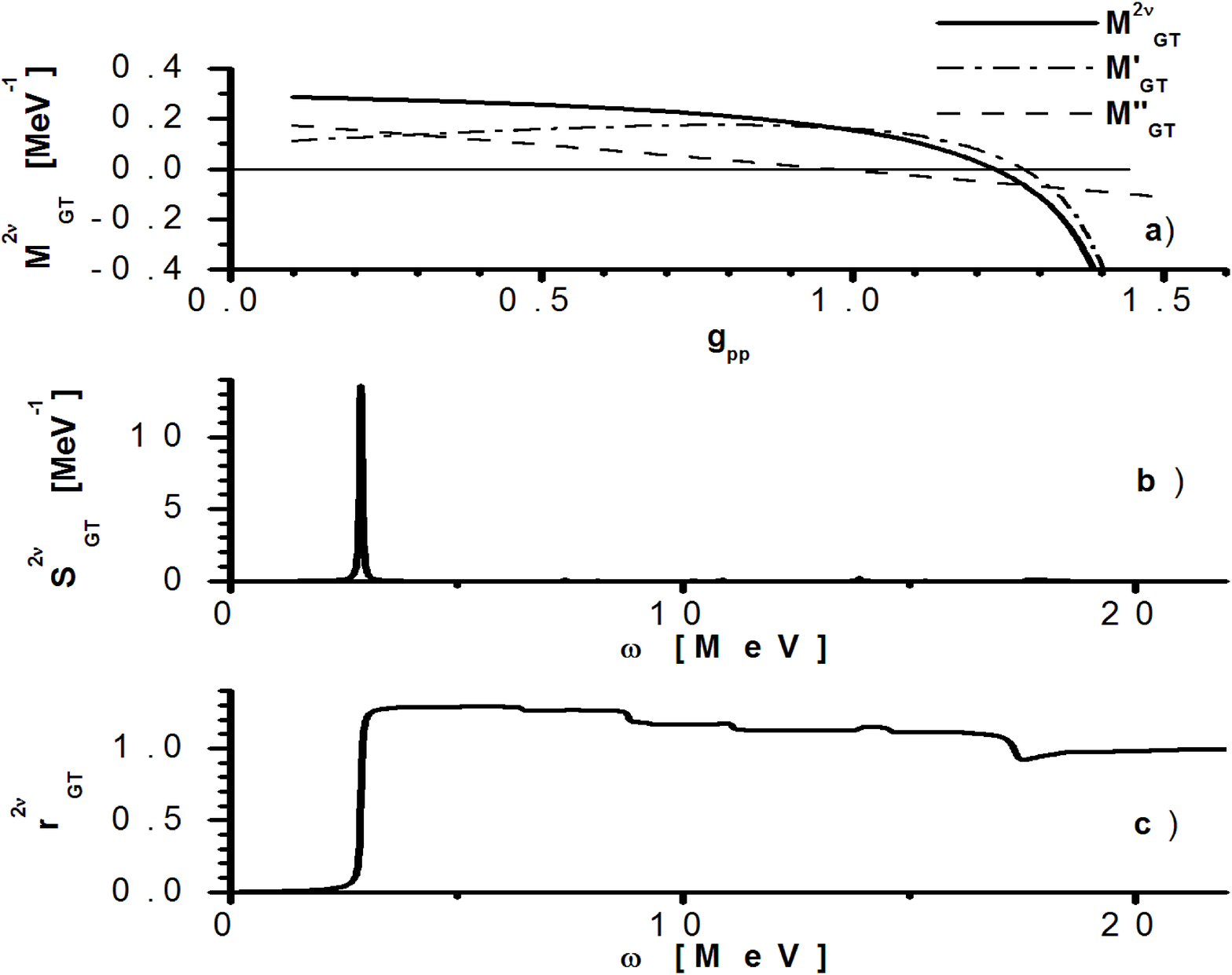}~~~~~\includegraphics[
width=8cm,height=14cm]{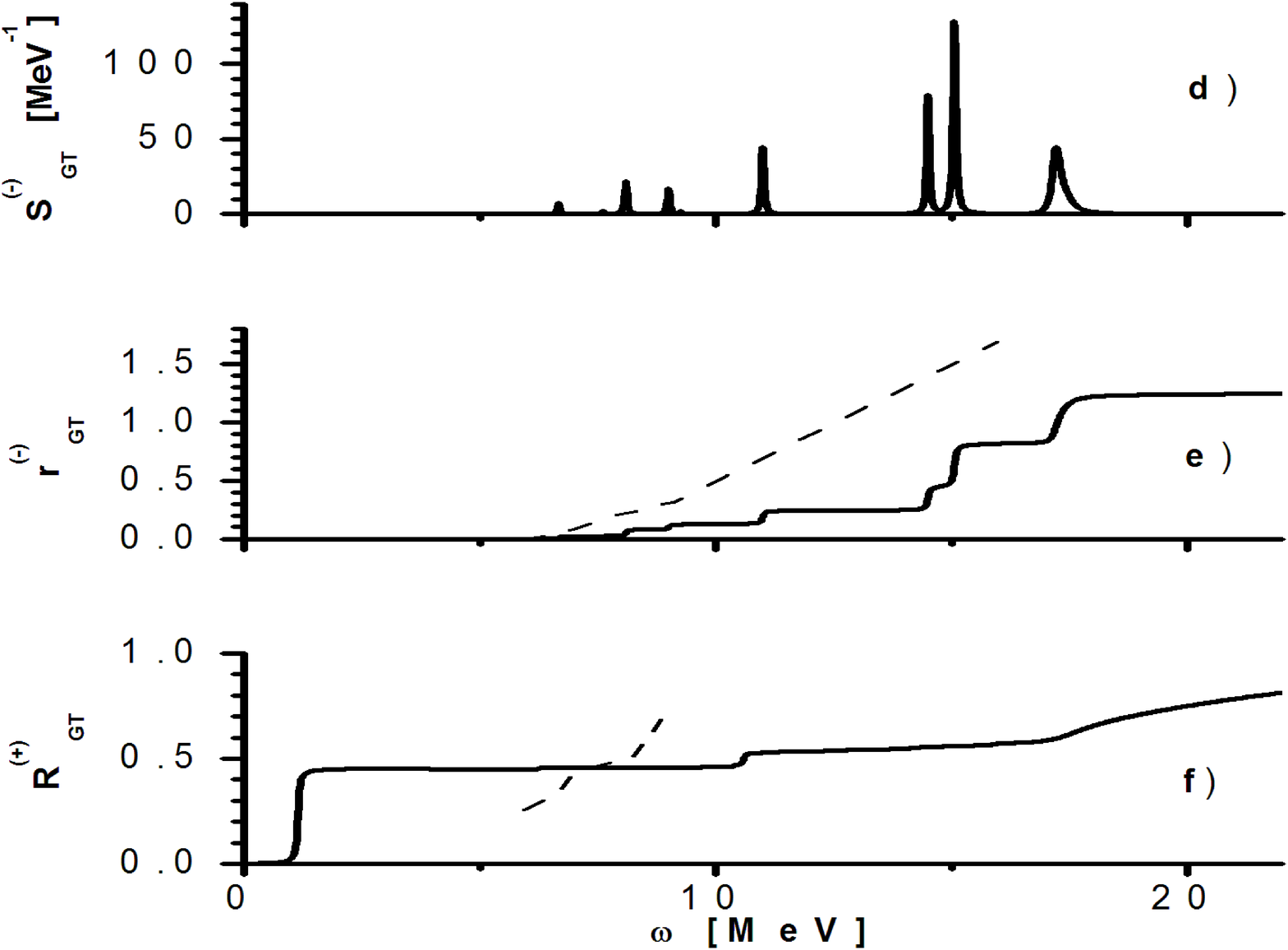}
\caption{ 
a)-c) the same as in Figs. 1.a)-1.c) but for the pair $^{116}$Cd-Sn; 
d) and e) the same as in Figs. 1.d) and 1.e) but for the pair $^{116}$Sn-Sb, 
the experimental data 
(related to the differential ($^3$He,t)-reaction cross section at $0^o$) 
are taken from Ref.~\cite{17}; 
f) the same as in Fig. 1.f) but for the pair $^{116}$Sn-In, the experimental data 
are taken from Ref.~\cite{3}.
}
\label{fig3}
\end{figure}

\begin{figure}[htb]
\includegraphics[
width=8cm,height=14cm]{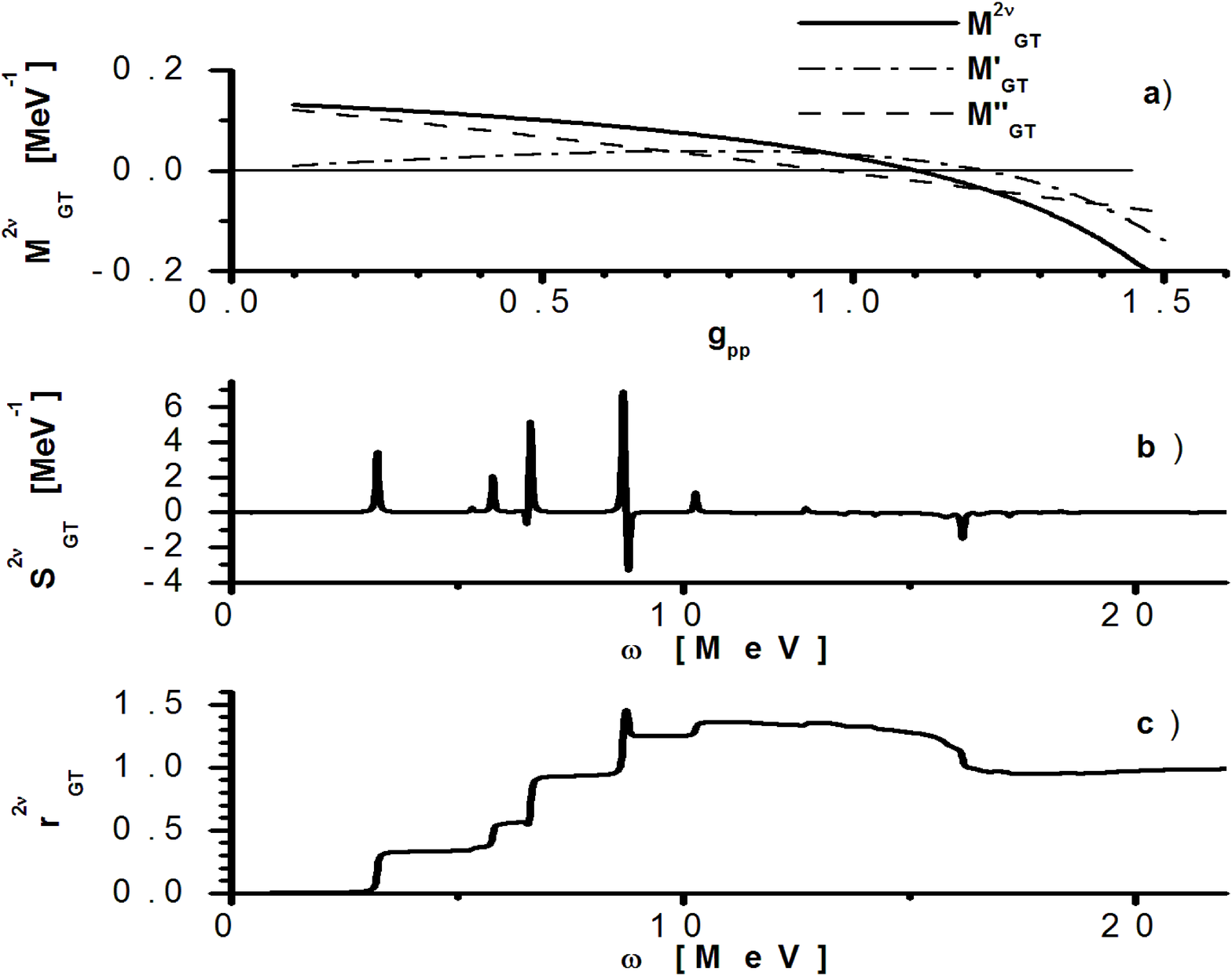}~~~~~\includegraphics[
width=8cm,height=14cm]{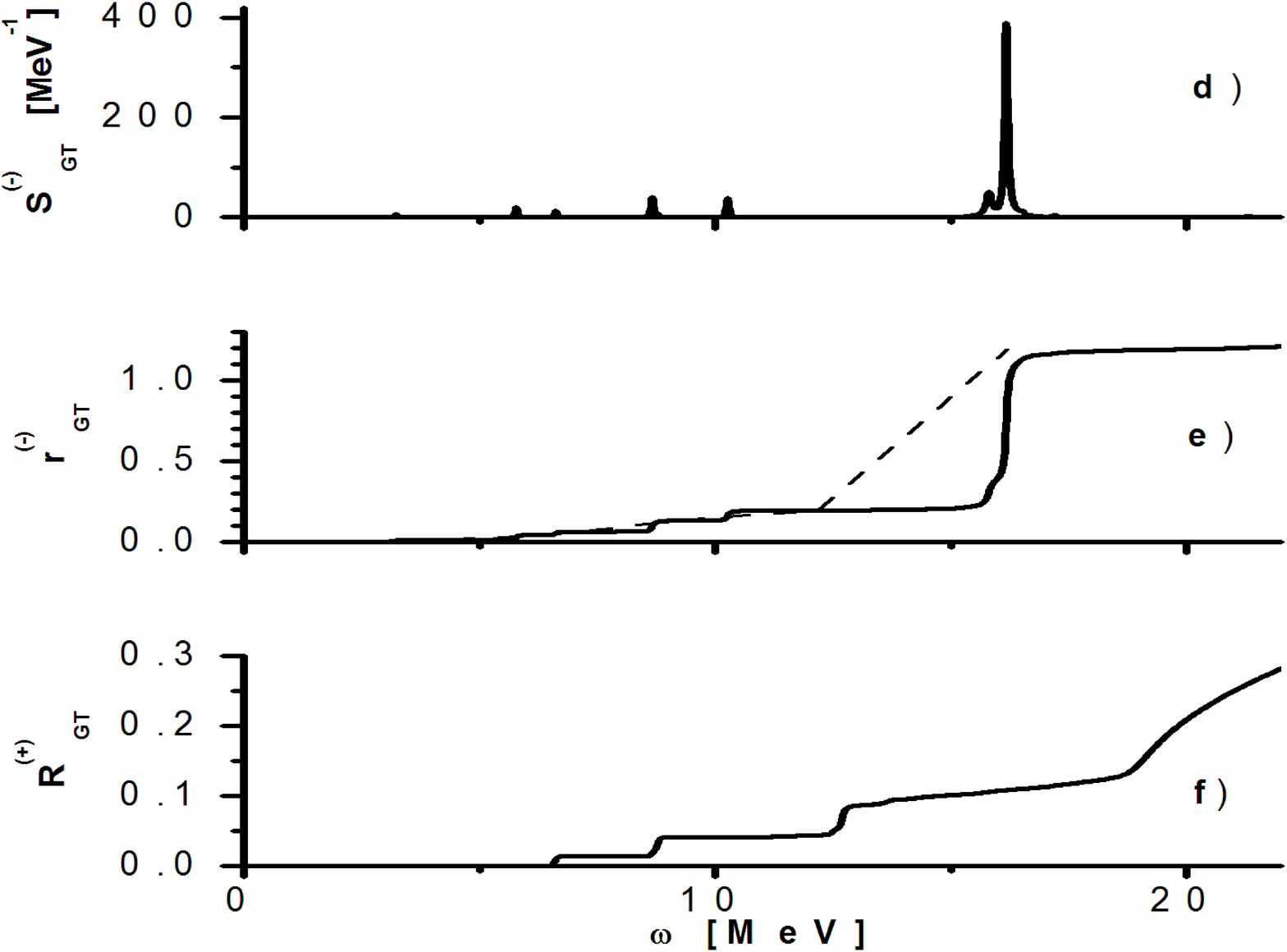}
\caption{ 
a)-c) the same as in Figs. 1.a)-1.c) but for the pair $^{130}$Te-Xe; 
d) and e) the same as in Figs. 1.d) and 1.e) but for the pair $^{130}$Te-I, 
the experimental data are taken from Ref.~\cite{15}; 
f) the same as in Fig. 1.f) but for the pair $^{130}$Xe-I.
}
\label{fig4}
\end{figure}

\end{document}